\documentclass[aps,superscriptaddress]{revtex4}
\usepackage{natbib}
\usepackage{ulem}
\usepackage{latexsym}
\usepackage{graphics}
\usepackage{amsfonts}
\usepackage{eurosym}
\usepackage{amsmath}
\usepackage{amssymb}
\usepackage{graphicx}
\usepackage[papersize={8.5in,11in}]{geometry}
\usepackage{hyperref}

\begin{document}
	\title{Optical Response and Drift Matrix of Quadratic Optomechanical System}
	\author{A. Kundu}
	\email{kunduaku339@gmail.com}
	\affiliation{Lovely Professional University, Phagwara 144411, Punjab, India.}
	\begin{abstract}
		Nonlinear interactions in optomechanical systems play a crucial role in many emerging number of interesting studies and phenomena such as existence of optomechanical chaos introduced by F. Monifi \textit{et al.} [\href{https://nature.com/articles/nphoton.2016.73}{\textit{Nature Photonics} \textbf{10}, 399–405 (2016)}] and optomechanical symmetry breaking proposed by Zhong-Peng Liu \textit{et al.}  [\href{https://journals.aps.org/prl/abstract/10.1103/PhysRevLett.117.110802}{\textit{Phys. Rev. Lett.}\textbf{117}, 110802 (2016)}]. In this article we have theoretically examined quadratically coupled optomechanical system containing two atomic levels. We have first studied the solution of various modes of the system at steady state and later we have observed the variation of \textit{Transmission Intensity} ($\mathcal{T}$) with several parameters of the system. Further we have extended our analyzation to find \textit{Drift matrix} of the quadratic optomechanical system and stability conditions by adiabetically eliminating atomic degree of freedom.\\\\
		\textbf{keywords:} Nonlinear interaction . Optomechanical system . Langevin equation . Transmission amplitude . Drift matrix . Adiabetic elimination
	\end{abstract}
	\maketitle
	\section{Introduction}
	Optomechanical coupling via radiation pressure is a
	very successful approach to prepare and manipulate quantum states of mechanical oscillators. A widely used cavity optomechanical system, is represented by a single mode Febry-P\'erot cavity with one movable end (i.e. less heavy mirror). The average position of the movable mirror is controlled by radiation pressure of light intensity inside cavity. This circulating intensity introduces an interaction between the cavity and mechanical degree of freedom. Coupling of optical and mechanical degrees of freedom influenced by externally applied radiation pressure has many applications from gravitational-wave detectors\cite{caves1,loudon2} to laser cooling\cite{hansch3,wineland4,chu5}. A large amount of studies of optomechanical systems deals with linear optomechanical coupling where, the cavity mode couples to the position of the movable mirror. Linear coupling has wide variety of applications such as, it is used for quantum ground state cooling\cite{arcizet6,wilson7,marquardt8,chan9,teufel10} of the mechanical mirror, entanglement between light and the mirror\cite{vitali11,palomaki12,aranya12.5}, electromagnetically induced transparency\cite{huang13,safavi14}, optomechanical induced transparency\cite{weis15}, squeezing\cite{akash15.5} and studies concerning normal-mode splitting\cite{dobrindt16,groblacher17}. In several works quadratic optomechanical coupling have also been considered where, the cavity mode is coupled to the square of position of mechanical operator. Quadratic optomechanical coupling has many applications such as it has been used to observe quantization in mechanical energy\cite{thompson18}, traditional two phonon laser cooling\cite{nunnenkamp19}, tunable slow light\cite{zhan20}, photon blockade\cite{liao21}, optomechanics at a single photon level\cite{liao22} etc. In this article we have considered the variation of \textit{transmission intensity} of an optically driven quadratic optomechanical system containing two atomic levels with various parameters of the following system. Previously, the variation of optical response (optical transmission) within two sideband limits\cite{kong22.5} of cavity detuning for a linearly coupled optomechanical system has been discussed analytically as well as numerically in Ref.\cite{farooq23} which we have reproduced on the way for quadratic optomechanical system.\\
	This work is organized as follows, in section (\ref{sec1}) we have briefly discussed the constriction of optomechanical system followed by the Hamiltonian in mode as well as quadrature representation. In section (\ref{sec2}) we have observed the dynamics of the system using \textit{quantum Langevin equation} which followed by a steady state treatment. Under the steady state conditions we have analytically found out the dependence of cavity mode on mechanical and atomic modes. A mathematical expression of \textit{transmission amplitude} and \textit{transmission intensity} has been obtained by using the results of section (\ref{sec2}) in (\ref{sec3}) followed by schematic plotting of \textit{transmission amplitude} with various parameters of the system. In the section (\ref{sec4}) we have constructed a drift matrix of the quadratically coupled optomechanical system by observing dynamics of the system and introducing infinitesimal \textit{quantum fluctuations} to the steady state values. Finally the article has been concluded in section (\ref{sec5}).
	\section{System Hamiltonian}\label{sec1}
	In our system, we have considered an optomechanical system containing two level electronic system. Our system consists of mode of single frequency of single Febry-P\'erot optical cavity and a single two level system inside the cavity. The arrangement of cavity is such that, it has one mirror which is heavier than the other and the heavier mirror is fixed in space hence, the lighter mirror can oscillate while exposed to radiation pressure with variable driven frequency ($\omega_D$). To find the dynamical as well as steady state solutions we have used quantum Langevin equations. The optomechanical system has been driven by an external coherent field, $\alpha_c=e^{-i\omega_Dt}$ of single mode driven frequency $\omega_D$. The smaller and less heavier mirror is oscillating at a single mode frequency $\omega_m$ under the radiation pressure and its damping rate denoted by $\Gamma$. Hence our system Hamiltonian will be divided into two parts, $\mathcal{H}_{om}$ representing the Hamiltonian of the optomechanical system and $\mathcal{H}_e$ represents the Hamiltonian of two level electronic system. The optomechanical hamiltonian of the system is given by,
	\begin{equation}
	\hat{\mathcal{H}}_{om}=\hbar\omega_a\hat{a}^{\dagger}\hat{a}+\hbar\omega_m\hat{b}^\dagger\hat{b}+\hbar g_{op}\hat{a}^\dagger\hat{a}\left(\hat{b}+\hat{b}^\dagger\right)^2\\
	-i\eta\hbar(\hat{a}^{\dagger}e^{-i\omega_Dt}-H.c.)\label{eq:om-hamiltonian}
	\end{equation}
	In Eq.(\ref{eq:om-hamiltonian}) the first and second term represents free energy of optical cavity and mechanical resonator with $\hat{a}^{\dagger}$($\hat{a}$) and $\hat{b}^{\dagger}$($\hat{b}$) are the creation(annihilation) operators of cavity and mechanical resonator. $g_{op}$ is the optomechanical coupling strength between cavity and mechanical mode(less heavier mirror) of the system.\\
	After inserting the two level system inside the cavity, the contribution of two-level system must be included via the Hamiltonian,
	\begin{equation}
	\hat{\mathcal{H}}_e=\dfrac{\hbar}{2}\omega_e\hat{\sigma}_z+\hbar g(\hat{\sigma}_+\hat{a}+\hat{\sigma}_-\hat{a}^{\dagger})\label{eq:electronic-hamiltonian} 
	\end{equation}
	where, $g$ is the coupling constant between the cavity field and two level system and, $\hat{\sigma}_+$, $\hat{\sigma}_-$ and, $\hat{\sigma}_z$ are the Pauli operators. For low excitation probability, the quantity, $\dfrac{1}{2}(\langle\hat{\sigma}_z\rangle+1)\ll1$, which allows us to apply the \textit{Holstein-Primakoff}\cite{holstein23.5} approximation in which atom is modeled by harmonic oscillator, so $\hat{\sigma}_-\rightarrow\hat{e}$, $\hat{\sigma}_+\rightarrow \hat{e}^{\dagger}$ and $\hat{\sigma}_z\approx2\hat{e}^{\dagger}\hat{e}$. Hence the atom Hamiltonian of our system,
	\begin{equation}
	\hat{\mathcal{H}}_e=\hbar\omega_e\hat{e}^{\dagger}\hat{e}+\hbar g(\hat{e}^{\dagger}\hat{a}+\hat{e}\hat{a}^{\dagger})\label{eq:electronic-hamiltonian-final}
	\end{equation}
	The total Hamiltonian of the system,
	\begin{equation}
	\hat{\mathcal{H}}=\hat{\mathcal{H}}_{om}+\hat{\mathcal{H}}_e\nonumber
	\end{equation}
	using Eq.(\ref{eq:om-hamiltonian}) and Eq.(\ref{eq:electronic-hamiltonian-final}) we get,
	\begin{equation}
	\hat{\mathcal{H}}=\hbar\omega_a\hat{a}^{\dagger}\hat{a}+\hbar\omega_m\hat{b}^\dagger\hat{b}+\hbar g_{op}\hat{a}^\dagger\hat{a}\left(\hat{b}+\hat{b}^\dagger\right)^2\\
	+\hbar\omega_e\hat{e}^{\dagger}\hat{e}+\hbar g(\hat{e}^{\dagger}\hat{a}+\hat{e}\hat{a}^{\dagger})-i\eta\hbar(\hat{a}^{\dagger}\hat{e}^{-i\omega_Dt}-H.c.)\label{eq:final-hamiltonian}
	\end{equation}
	\subsection{Quadrature Representation}
	We may define our optomechanical Hamiltonian (\ref{eq:om-hamiltonian}) in terms of quadrature of the mechanical mode i.e. $\hat{b}=\dfrac{\hat{Q}+i\hat{P}}{\sqrt{2}}$ and $\hat{b}^\dagger=\dfrac{\hat{Q}-i\hat{P}}{\sqrt{2}}$ as,
	\begin{equation}
	\hat{\mathcal{H}}_{om}=\hbar\omega_a\hat{a}^{\dagger}\hat{a}+\hbar\dfrac{\omega_m}{2}\left(\hat{Q}^2+\hat{P}^2-1\right)+2i\hbar g_{op}\hat{a}^\dagger\hat{a}\hat{Q}^2\\
	-i\eta\hbar(\hat{a}^{\dagger}\hat{e}^{-i\omega_Dt}-H.c.)\label{eq:om-hamiltonian-quadrature}
	\end{equation}
	\section{Dynamics of the System}\label{sec2}
	To get valuable information of the optomechanical system defined by Eq. (\ref{eq:final-hamiltonian}) we need to observe the dynamics of the system. For this purpose we used \textit{Quantum Langevin Equations} given by,
	\begin{subequations}
		\begin{align}
		&\dfrac{d\hat{a}}{dt}=\dfrac{i}{\hbar}[\hat{\mathcal{H}},\hat{a}]-\dfrac{\epsilon}{2}\hat{a}+\sqrt{2\epsilon}\hat{a}_{in}\label{eq:cavity-langevin}\\
		&\dfrac{d\hat{b}}{dt}=\dfrac{i}{\hbar}[\hat{\mathcal{H}},\hat{b}]-\dfrac{\Gamma}{2}\hat{b}+\sqrt{2\Gamma}\hat{b}_{in}\\
		&\dfrac{d\hat{e}}{dt}=\dfrac{i}{\hbar}[\hat{\mathcal{H}},\hat{e}]-\dfrac{\gamma}{2}\hat{e}+\sqrt{2\gamma}\hat{e}_{in}\label{eq:atom-langevin}
		\end{align}
	\end{subequations}
	where, $\epsilon$, $\Gamma$ and $\gamma$ are cavity, mirror and atomic mode damping. Now by expanding Eq.(\ref{eq:cavity-langevin})-(\ref{eq:atom-langevin}) we get,
	\begin{subequations}
		\begin{equation}
		\dfrac{d\hat{a}}{dt}=-i\omega_a\hat{a}-ig\hat{e}-ig_{op}\hat{a}\left(\hat{b}+\hat{b}^\dagger\right)^2-\eta e^{-i\omega_Dt}-\dfrac{\epsilon}{2}\hat{a}+\sqrt{2\epsilon}\hat{a}_{in}\label{eq:cavity-langevin-final}
		\end{equation}
		\begin{align}
		&\dfrac{d\hat{b}}{dt}=-i\omega_m\hat{b}-2ig_{op}\hat{a}^\dagger\hat{a}\left(\hat{b}+\hat{b}^\dagger\right)-\dfrac{\Gamma}{2}\hat{b}+\sqrt{2\Gamma}\hat{b}_{in}\label{eq:mechanical-langevin-final}\\\nonumber\\
		&\dfrac{d\hat{e}}{dt}=-i\omega_e\hat{e}-ig\hat{a}-\dfrac{\gamma}{2}\hat{e}+\sqrt{2\gamma}\hat{e}_{in}\label{eq:atom-langevin-final}
		\end{align}
		The nonlinear terms in Eqs.(\ref{eq:cavity-langevin-final}) and (\ref{eq:mechanical-langevin-final}) are arising due to the coupling between cavity and the quadratically moving mirror. Simplification of these type of equations are difficult to handle due to the complexity of calculations but with the help of quadratic representation given in Eq.(\ref{eq:om-hamiltonian-quadrature}) it can be easily analysed. To simplify the Eqs.(\ref{eq:cavity-langevin-final})-(\ref{eq:atom-langevin-final}) we use \textit{Rotating Wave Approximation} (RWA) where we replace $\hat{a}\rightarrow \hat{a}[exp(-i\omega_Dt)]$
		and, $\hat{e}\rightarrow\hat{e}[exp(-i\omega_Dt)]$ and arrive at,
	\end{subequations}
	\begin{subequations}
		\begin{align}
		&\dfrac{d\hat{a}}{dt}=-i\Delta_a\hat{a}-ig\hat{e}-ig_{op}\hat{a}\left(\hat{b}+\hat{b}^\dagger\right)^2-\eta-\dfrac{\epsilon}{2}\hat{a}+\sqrt{2\epsilon}\hat{a}_{in}\label{eq:cavity-langevin-final-rwa}\\
		&\dfrac{d\hat{b}}{dt}=-i\omega_m\hat{b}-2ig_{op}\hat{a}^\dagger\hat{a}\left(\hat{b}+\hat{b}^\dagger\right)-\dfrac{\Gamma}{2}\hat{b}+\sqrt{2\Gamma}\hat{b}_{in}\label{eq:mechanical-langevin-final-rwa}\\
		&\dfrac{d\hat{e}}{dt}=-i\Delta_e\hat{e}-ig\hat{a}-\dfrac{\gamma}{2}\hat{e}+\sqrt{2\gamma}\hat{e}_{in}\label{eq:atom-langevin-final-rwa}
		\end{align}
		where, $\Delta_i=\omega_i-\omega_D$, $i=a \ \text{and} \ e$ representing cavity and atomic level detuning. The expectation values of the system's operators can be used as observables to get proper information of the dynamical behaviour of the system hence, from Eqs.(\ref{eq:cavity-langevin-final-rwa})-(\ref{eq:mechanical-langevin-final-rwa}) we get,
		\begin{align}
		&\dfrac{d\hat{\langle a\rangle}}{dt}=-i\Delta_a\langle\hat{a}\rangle-ig\langle\hat{e}\rangle-ig_{op}\langle\hat{a}\rangle\left(\hat{b}+\hat{b}^\dagger\right)^2-\eta-\dfrac{\epsilon}{2}\langle\hat{a}\rangle\label{eq:cavity-langevin-final-rwa-expectationvalue}\\
		&\dfrac{d\langle\hat{b}\rangle}{dt}=-i\omega_m\langle\hat{b}\rangle-2ig_{op}\langle\hat{a}^\dagger\hat{a}\rangle\left(\hat{b}+\hat{b}^\dagger\right)-\dfrac{\Gamma}{2}\langle\hat{b}\rangle\label{eq:mechanical-langevin-final-rwa-expectationvalue}\\
		&\dfrac{d\langle\hat{e}\rangle}{dt}=-i\Delta_e\langle\hat{e}\rangle-ig\langle\hat{a}\rangle-\dfrac{\gamma}{2}\langle\hat{e}\rangle\label{eq:atom-langevin-final-rwa-expectationvalue}
		\end{align}
		where we have consider the expectation values of the noises are null i.e. $\langle \hat{a}_{in}\rangle=\langle \hat{b}_{in}\rangle=\langle\hat{e}_{in}\rangle=0$
	\end{subequations}
	\subsection{Steady State Dynamics}
	In this section we are about to study the steady state dynamics of the system which can be achieved by equating the \textit{RHS} of Eqs.(\ref{eq:cavity-langevin-final-rwa-expectationvalue})-(\ref{eq:atom-langevin-final-rwa-expectationvalue}) to zero,
	\begin{subequations}
		\begin{align}
		&0=-i\Delta_a\langle\hat{a}\rangle-ig\langle\hat{e}\rangle-ig_{op}\langle\hat{a}\rangle\left(\hat{b}+\hat{b}^\dagger\right)^2-\eta-\dfrac{\epsilon}{2}\langle\hat{a}\rangle\\
		&0=-i\omega_m\langle\hat{b}\rangle-2ig_{op}\langle\hat{a}^\dagger\hat{a}\rangle\left(\hat{b}+\hat{b}^\dagger\right)-\dfrac{\Gamma}{2}\langle\hat{b}\rangle\\
		&0=-i\Delta_e\langle\hat{e}\rangle-ig\langle\hat{a}\rangle-\dfrac{\gamma}{2}\langle\hat{e}\rangle
		\end{align}
	\end{subequations}
	Further analyzation leads to,
	\begin{subequations}
		\begin{align}
		&0=-\left(\dfrac{\epsilon}{2}+i\Delta_a\right)\langle\hat{a}\rangle-ig\langle\hat{e}\rangle-ig_{op}\langle\hat{a}\rangle\left(\hat{b}+\hat{b}^\dagger\right)^2-\eta\\
		&0=-\left(\dfrac{\Gamma}{2}+i\omega_m\right)\langle\hat{b}\rangle-2ig_{op}\langle\hat{a}^\dagger\hat{a}\rangle\left(\hat{b}+\hat{b}^\dagger\right)\\
		&0=-\left(\dfrac{\gamma}{2}+i\Delta_e\right)\langle\hat{e}\rangle-ig\langle\hat{a}\rangle
		\end{align}
	\end{subequations}
	by simplifying the above expressions we get,
	\begin{align}
	&\langle\hat{e}\rangle=\dfrac{-ig\langle \hat{a}\rangle}{\left(\dfrac{\gamma}{2}+i\Delta_e\right)}\label{eq:expression-e-steadystate}\\
	&\langle\hat{b}\rangle=\dfrac{-2ig_{op}\langle\hat{a}^{\dagger}\hat{a}\rangle(\hat{b}+\hat{b}^{\dagger})}{\left(\dfrac{\Gamma}{2}+i\omega_m\right)}\label{eq:expression-b-steadystate}\\
	&-\left[\left(\dfrac{\epsilon}{2}+i\Delta_a\right)+ig_{op}(\hat{b}+\hat{b}^{\dagger})^2\right]\langle\hat{a}\rangle-ig\langle\hat{e}\rangle=\eta\label{eq:expression-a-steadystate}
	\end{align}
	Using Eq.(\ref{eq:expression-e-steadystate}) on Eq.(\ref{eq:expression-a-steadystate}) we finally get,
	\begin{equation}
	\langle\hat{a}\rangle=-\dfrac{\eta}{\left(\dfrac{\epsilon}{2}+i\Delta\right)+\dfrac{g^2}{\left(\dfrac{\gamma}{2}+i\Delta_e\right)}}\label{eq:expression-a-steadystate-final}	
	\end{equation}
	where, $\Delta=\Delta_a+g_{op}(\hat{b}+\hat{b}^{\dagger})^2$
	\section{Transmission Amplitude and Intensity}\label{sec3}
	To observe the optical response of the cavity with the environment we need to focus on the transmission intensity of the system which can be evaluated after calculating complex transmission amplitude. By using input-output formalism we found the transmission amplitude as\cite{farooq23,collett24},
	\begin{equation}
	\mathcal{A}_T=-\eta\langle\hat{a}\rangle
	\end{equation}
	by using Eq.(\ref{eq:expression-a-steadystate-final}) we get,
	\begin{equation}
	\mathcal{A}_T=\dfrac{\eta^2}{\left(\dfrac{\epsilon}{2}+i\Delta\right)+\dfrac{g^2}{\left(\dfrac{\gamma}{2}+i\Delta_e\right)}}\label{eq:transmission-amplitude}
	\end{equation}
	The \textit{Transmission Intensity} can be evaluated as, $\mathcal{T}=|\mathcal{A}_T|^2$.
	\begin{equation}
	\mathcal{T}=\dfrac{\eta^4}{\left(\dfrac{\epsilon^2}{4}+\Delta^2\right)+\dfrac{g^4+(\dfrac{\epsilon\gamma}{2}-2\Delta\Delta_e)g^2}{\left(\dfrac{\gamma^2}{4}+\Delta_e^2\right)}}\label{eq:transmission-amplitde}
	\end{equation}
	\subsection{Numerical Plotting}
	\begin{figure}
		\includegraphics[width=120mm]{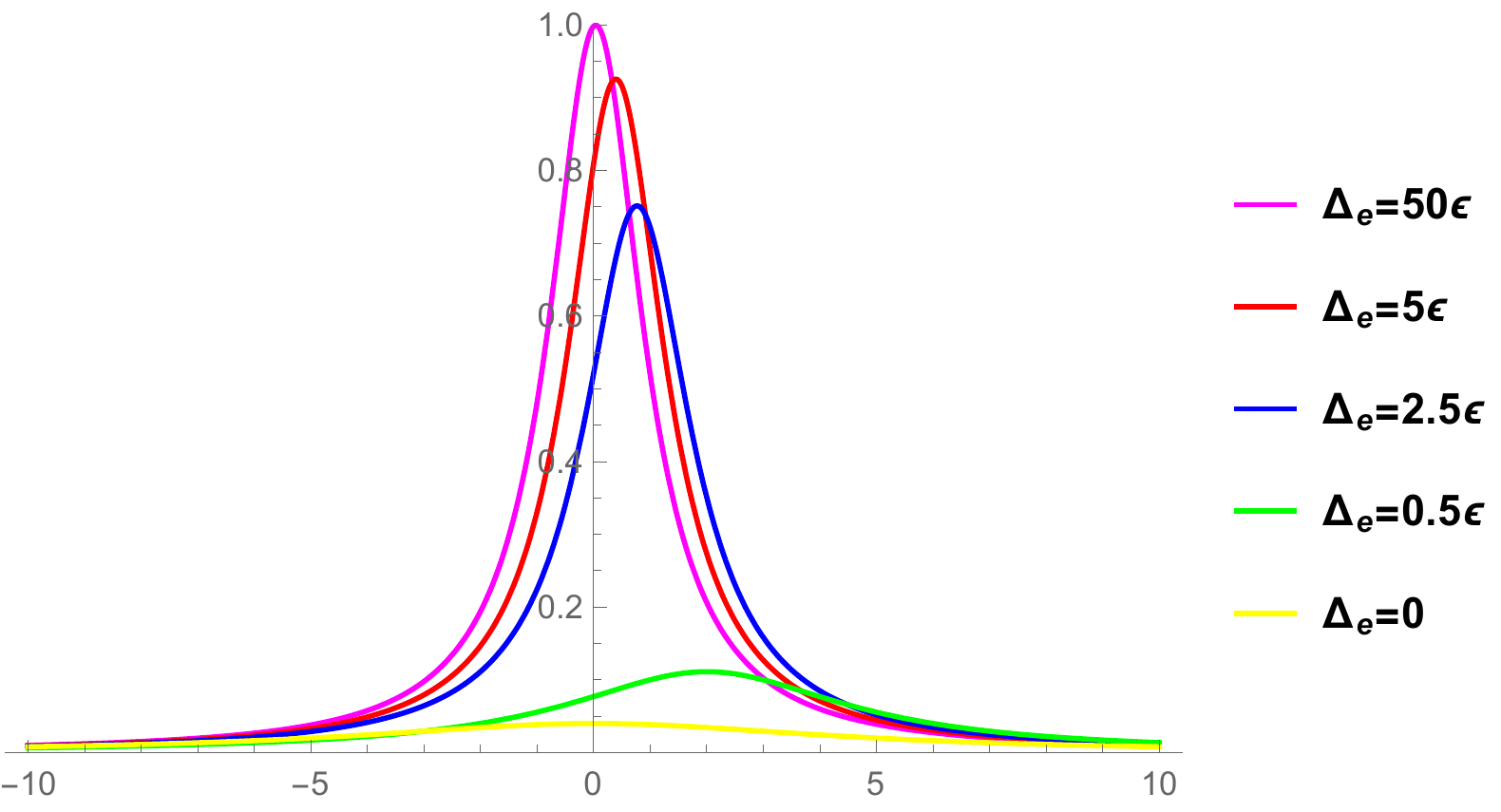}
		\caption{The variation of $\mathcal{T}$ with $\Delta$ for $(1)$ $\Delta_e=0$, $(2)$ $\Delta_e=0.5\epsilon$, $(3)$ $\Delta_e=2.5\epsilon$, $(4)$ $\Delta_e=5\epsilon$ and $(5)$ $\Delta_e=50\epsilon$. It is observed as, the value of $\Delta_e=\omega_e-\omega_D$ increases, mismatch between the frequencies of incident optical pumping and two level system atom raises; which leads to an enhancement in \textit{Transmission Amplitude}.}
		\label{fig:variation-with-delta}
	\end{figure}
	\begin{figure}
		\includegraphics[width=120mm]{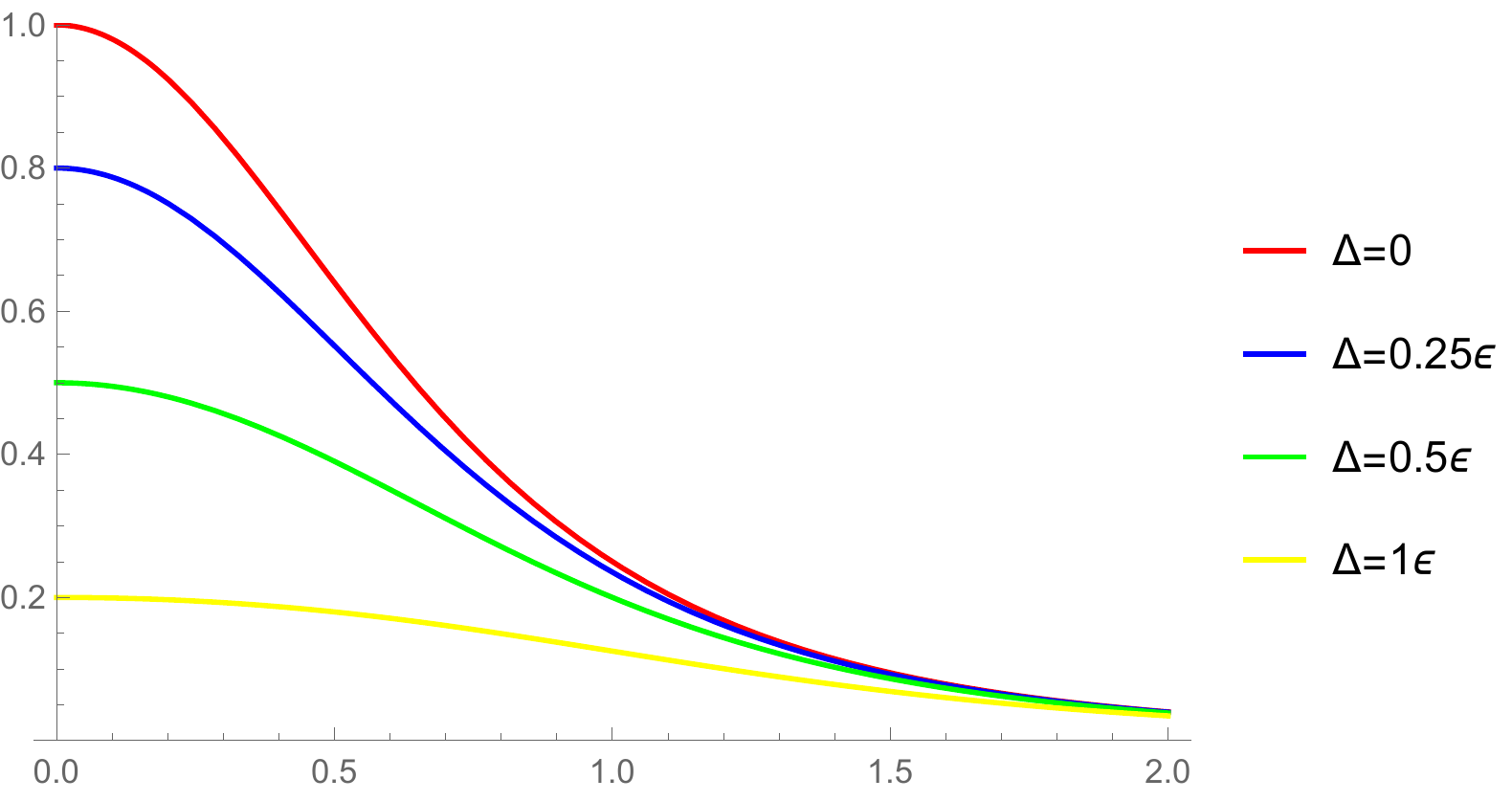}
		\caption{Variation of $\mathcal{T}$ with $g$ for $(1)$ $\Delta=0$, $(2)$ $\Delta=0.25\epsilon$, $(3)$ $\Delta=0.5\epsilon$ and, $(4)$ $\Delta=1\epsilon$. It is observed from the plot that, at $\Delta=0$, $\mathcal{T}$ shows maximum value with no coupling between the atomic levels and optical pumping and, as the coupling increases $\mathcal{T}$ degrades rapidly.}
		\label{fig:variation-with-g}
	\end{figure}
\begin{figure}
		\includegraphics[width=120mm]{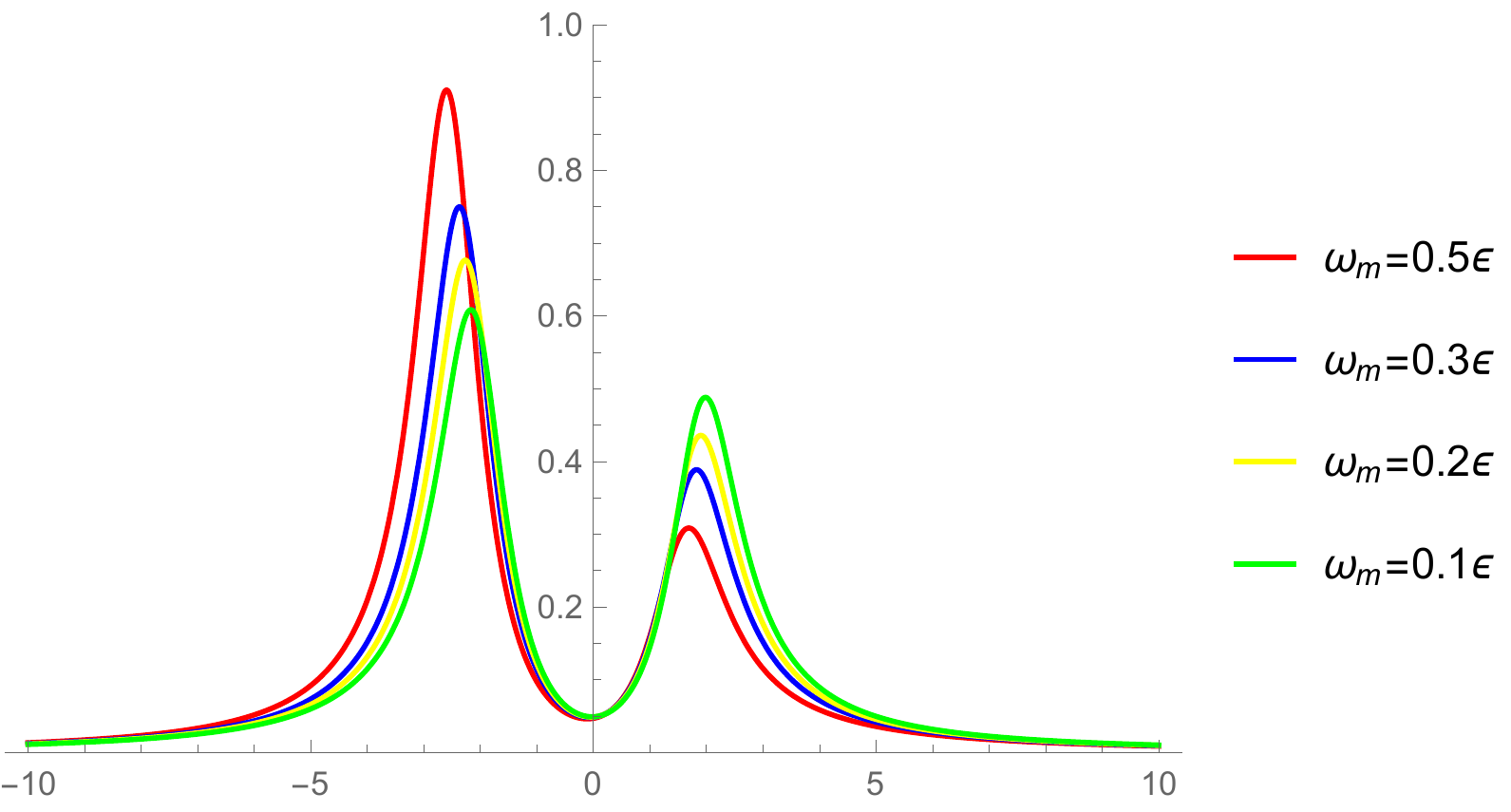}
		\caption{Optical \textit{transmission intensity} $\mathcal{T}$ as a function of cavity detuning $\Delta_a$. For (1) $\omega_m=0.1\epsilon$, (2) $\omega_m=0.2\epsilon$, (3) $\omega_m=0.3\epsilon$ and, (4) $\omega_m=0.5\epsilon$.}
		\label{fig:transmission-with-cavity-detuning}
	\end{figure}

	In the numerical plottings all frequencies are in the unit of $\epsilon$. Fig. [\ref{fig:variation-with-delta}] shows the variation of $\mathcal{T}$ with $\Delta$ for $\epsilon=2$, $\gamma=\epsilon$, $\eta=\dfrac{\epsilon}{2}$ (in the unit of $\epsilon$) where, we can observe for $\Delta_e=0$, the $\mathcal{T}$ shows maximum value $0.04$ at $\Delta=0$ hence, under this specific condition $\Delta=0$ is the resonance position. For, $\Delta_e=1$ the $\mathcal{T}$ shows maximum value of $\approx0.112$ at $\Delta\approx1.95$ hence, $\Delta=1.95$ is the resonance position. For, $\Delta_e=5$ the $\mathcal{T}$ shows maximum value of $\approx0.76$ at $\approx\Delta=0.75$ hence, $\Delta=0.75$ is the resonance position. For, $\Delta_e=10$ the $\mathcal{T}$ shows maximum value of $\approx0.93$ at $\approx\Delta=0.4$ hence, $\Delta=0.4$ is the resonance position. Finally, for $\Delta_e=100$ the $\mathcal{T}$ shows maximum value of $1.0$ at $\Delta=0$ hence, $\Delta=0$ is the resonance position again. So, it can be observed form the plot that as, we increase $\Delta_e$ from $0$ to $100$ the resonance position varies from $0$ to $\approx2$ and gradually comes back to $0$ near $\Delta_e\approx100$. It has been observed that for a variation of $\Delta_e$ from $0$ to $5$; $\mathcal{T}$ varies rapidly from $0.04$ to $0.76$ but this variation rapidity drastically reduces when we vary $\Delta_e$ from $5$ to $100$.\\
	In Fig. [\ref{fig:variation-with-g}] we have plotted the variation of \textit{Transmission Intensity} with coupling coefficient $g$ for $(1)$ $\Delta=0$, $(2)$ $\Delta=0.25\epsilon$, $(3)$ $\Delta=0.5\epsilon$ and, $(4)$ $\Delta=1\epsilon$ by considering $\epsilon=2$, $\gamma=\epsilon$, $\eta=\dfrac{\epsilon}{2}$ and, $\Delta_e=0$. The figure shows variation of $\mathcal{T}$ with coupling between the cavity photons and two level system $g$ and it can be observed that, as the coupling varies from $0$ to $0.2$, $\mathcal{T}$ remains constant and, an increment in $g$ from $0.3$ to $1.0$ rapidly decreases $\mathcal{T}$ and the it gradually tends to saturate. Further increment in $g$ above $1.5$ puts $\mathcal{T}$ to saturation at a value $\approx0.05$. But at the same time an increase in $\Delta$ from $0$ to $2$ reduces $\mathcal{T}$ at position $g=0$ i.e. the maximum achievable value of $\mathcal{T}$ for a particular value of $\Delta$.\\
	In Fig.[\ref{fig:transmission-with-cavity-detuning}] we have reproduced the results demonstrated in Ref.\cite{farooq23} (see. Fig.[3] of Ref.\cite{farooq23}) by plotting transmission intensity $\mathcal{T}$ with the optically driven field frequency $\omega_D$ for \textit{Upper Sideband} of cavity detuning i.e. $\Delta=\Delta_a+\omega_m$ when $\omega_a=\omega_e=0$ for, $\eta=\dfrac{\epsilon}{2}$, $\epsilon=2$, $g=\epsilon$ and $\gamma=\epsilon$ for (1) $\omega_m=0.1\epsilon$, (2) $\omega_m=0.2\epsilon$, (3) $\omega_m=0.3\epsilon$ and, (4) $\omega_m=0.5\epsilon$. $\Delta=\Delta_a+g_{op}(\hat{b}+\hat{b}^{\dagger})^2=\Delta_a\pm\omega_m$ which describes \textit{Upper} and \textit{Lower Sideband} of \textit{cavity detuning}. Now the distinguishable factor of our outcome with the previous results described as follows,\\
	when, $g_{op}>0$ we can only consider the \textit{Upper sideband} of cavity detuning i.e. $(\Delta_a+\omega_m)$ which makes $g_{op}(\hat{b}+\hat{b}^\dagger)^2=\omega_m$ or, $2\hat{Q}^2=\dfrac{g_{op}}{\omega_m}$ and the mirror displacement $\hat{Q}=\pm\sqrt{\dfrac{g_{op}}{2\omega_m}}$ which completely describes the to and fro motion of the mechanical mirror due to the impact of radiation pressure but, for the \textit{Lower Sideband} i.e. ($\Delta_a-\omega_m$) and, $-g_{op}(\hat{b}+\hat{b}^\dagger)=\omega_m$ mirror displacement $\hat{Q}$ represented by a completely \textit{imaginary} entity which is \textit{impossible} as $\hat{Q}$ is real.
	\section{Drift Matrix}\label{sec4}
	In this section we will construct the \textit{Drift Matrix} for our system by considering optical pumping frequency $\omega_D\gg\omega_e$ which allows us to adiabetically eliminate the ionic degree of freedom i.e. $\mathcal{H}_e$ can be neglected. Hence our system Hamiltonian under \textit{RWA} takes the form,
	\begin{equation}
	\mathcal{H}=\hbar\Delta_a\hat{a}^{\dagger}\hat{a}+\hbar\omega_m\hat{b}^\dagger\hat{b}+\hbar g_{op}\hat{a}^\dagger\hat{a}\left(\hat{b}+\hat{b}^\dagger\right)^2\\
	-i\eta\hbar(\hat{a}^{\dagger}+\hat{a})\label{eq:om-hamiltonian-adiabetic}
	\end{equation}
	for the sake of simplicity, we will proceed with the quadrature representation of the Hamiltonian as,
	\begin{equation}
	\mathcal{H}=\hbar\Delta_a\hat{a}^{\dagger}\hat{a}+\hbar\dfrac{\omega_m}{2}\left(\hat{Q}^2+\hat{P}^2-1\right)+2i\hbar g_{op}\hat{a}^\dagger\hat{a}\hat{Q}^2\\
	-i\eta\hbar(\hat{a}^{\dagger}+\hat{a})\label{eq:om-hamiltonian-quadrature-adiabetic}
	\end{equation}
	\subsection{Dynamics of the System}
	The evolution of Hamiltonian given by Eq.(\ref{eq:om-hamiltonian-quadrature-adiabetic}) is given by the set of evolution equations obtained using quantum Langevin equation,
	\begin{align}
	&\dfrac{d\hat{a}}{dt}=-i\Delta_a\hat{a}-2ig_{op}\hat{a}\hat{Q}^2-\dfrac{\epsilon}{2}\hat{a}-\eta+\sqrt{2\epsilon}\hat{a}_{in}\label{eq:expression-a-adiabetic}\\
	&\dfrac{d\hat{P}}{dt}=-\omega_m\hat{Q}-4g_{op}\hat{a}^{\dagger}\hat{a}\hat{Q}-\dfrac{\Gamma}{2}\hat{P}+\sqrt{2\Gamma}\hat{P}_{in}\\
	&\dfrac{d\hat{Q}}{dt}=\omega_m\hat{P}+\sqrt{2\Gamma}\hat{Q}_{in}\label{eq:expression-P-adiabetic}
	\end{align}
	where, $\hat{P}_{in}$ and $\hat{Q}_{in}$ represents the quadrature input noise of mechanical oscillator.
	\subsection{Quantum Fluctuations}
	As the fluctuations in a quantum system is relatively negligible then the steady state values hence we can neglect the nonlinear terms arises due to the fluctuations i.e. $\hat{a}\rightarrow\hat{a}+\lambda\delta\hat{a}$, $\hat{P}\rightarrow\hat{P}+\lambda\delta\hat{P}$ and, $\hat{Q}=\hat{Q}+\lambda\delta\hat{Q}$ where we neglect nonlinear terms in $\lambda$,
	\begin{equation}
	\dfrac{d}{dt}\delta\hat{a}=-i\Delta_a\delta\hat{a}-4ig_{op}\hat{a}_s\hat{Q}_s\delta\hat{Q}-2ig_{op}\hat{Q}^2_s\delta\hat{a}-\dfrac{\epsilon}{2}\delta\hat{a}+\sqrt{2\epsilon}\delta\hat{a}_{in}
	\end{equation}
	\begin{equation}
	\dfrac{d}{dt}\delta\hat{P}=-\left(\omega_m+4g_{op}|a_{s}|^2\right)\delta\hat{Q}+\dfrac{\Gamma}{2}\delta\hat{P}-4g_{op}a_sQ_s\left(\dfrac{\hat{a}_s+\hat{a}_s^*}{\sqrt{2}}\right
	)+\sqrt{2\Gamma}\delta\hat{P}_{in}
	\end{equation}
	\begin{equation}
	\dfrac{d}{dt}\delta\hat{Q}=\omega_m\delta\hat{P}+\sqrt{2\Gamma}\delta\hat{Q}_{in}	
	\end{equation}
	Now using the expression, $\delta\hat{a}=\dfrac{\delta\hat{X}_a+i\delta\hat{P}_a}{\sqrt{2}}$; $\delta\hat{a}^\dagger=\dfrac{\delta\hat{X}_a-i\delta\hat{P}_a}{\sqrt{2}}$ and, $\dfrac{\hat{a}_s+\hat{a}_s^*}{\sqrt{2}}=X_s$, $\dfrac{\hat{a}_s-\hat{a}_s^*}{i\sqrt{2}}=P_s$
	we get,
	\begin{subequations}
		\begin{equation}
		\dfrac{d}{dt}\delta\hat{X}_a=\left(\Delta_a+2g_{op}Q_s^2\right)\delta\hat{P}_a+4g_{op}Q_sP_s\delta\hat{Q}-\dfrac{\epsilon}{2}\delta\hat{X}_a+\sqrt{2\epsilon}\delta\hat{X}^{in}_{a}
		\end{equation}
		\begin{equation}
		\dfrac{d}{dt}\delta\hat{P}_a=-\left(\Delta_a+2g_{op}Q_s^2\right)\delta\hat{X}_a-4g_{op}X_sQ_s\delta\hat{Q}-\dfrac{\epsilon}{2}\delta\hat{P}_a+\sqrt{2\epsilon}\delta\hat{P}^{in}_{a}
		\end{equation}
		\begin{equation}
		\dfrac{d}{dt}\delta\hat{Q}=\omega_m\delta\hat{P}+\sqrt{2\Gamma}\delta\hat{Q}_{in}	
		\end{equation}
		\begin{equation}
		\dfrac{d}{dt}\delta\hat{P}=-\left(\omega_m+4g_{op}|a_{s}|^2\right)\delta\hat{Q}-\dfrac{\Gamma}{2}\delta\hat{P}-4g_{op}X_sQ_s\delta{X}_a+\sqrt{2\Gamma}\delta\hat{P}_{in}
		\end{equation}
	\end{subequations}
	The \textit{Drift Matrix} written as,
	\[
	M=
	\left( {\begin{array}{cccc}
		-\dfrac{\epsilon}{2} & \tilde{\Delta}_a & GP_s & 0\\
		-\tilde{\Delta}_a & -\dfrac{\epsilon}{2} & -GX_s & 0\\
		0 & 0 & 0 & \omega_m\\
		-GX_s & 0 & -\tilde{\omega}_m & -\dfrac{\Gamma}{2}
		\end{array} } \right)
	\]
	where, 
	\begin{align}
	&\tilde{\Delta}_a=\Delta_a+2g_{op}Q_s^2\nonumber\\ &\tilde{\omega}_m=\omega_m+4g_{op}|a_s|^2\nonumber\\
	&G=4g_{op}Q_s\nonumber
	\end{align}
	and, the evolution equation of the optomechanical system due to quantum flactuations,
	\begin{equation}
	\dot{X}=MX+N\label{eq:evolution-eqn}
	\end{equation}
	where, $X^T=\left({\begin{array}{cccc}
		\delta X_a & \delta P_a & \delta Q & \delta P
		\end{array}}\right)$ representing the infinitesimal change in the system parameters and, $N^T=\left({\begin{array}{cccc}
		\sqrt{2\epsilon}\delta X_a^{in} & \sqrt{2\epsilon}\delta P_a^{in} & \sqrt{2\Gamma}\delta Q_{in} & \sqrt{2\Gamma}\delta P_{in}
		\end{array}}\right)$ the noise matrix of the system.
	\subsection{Stability Conditions}
	The stability conditions can be deduced by applying
	the Routh-Hurwitz\cite{routh25} criteria by finding the $n^{\text{th}}$ polynomial equation of eigenvalues. And, the conditions for stability are,
	\begin{align}
	&l_1=\left(\dfrac{\epsilon^2}{4}+\tilde{\Delta}_a^2\right)+\dfrac{\epsilon\Gamma}{2}+\omega_m\tilde{\omega}_m>0\\
	&l_2=\epsilon\omega_m\tilde{\omega}_m+G^2X_sP_s\omega_m+\left(\dfrac{\epsilon^2}{4}+\tilde{\Delta}_a^2\right)\dfrac{\Gamma}{2}>0\\
	&l_3=\omega_m\tilde{\omega}_m\left(\dfrac{\epsilon^2}{4}+\tilde{\Delta}_a^2\right)+\dfrac{G^2\epsilon}{2}P_sX_a\omega_m-G^2X_S^2\omega_m\tilde{\omega}_m>0\\
	&\left(\epsilon+\dfrac{\Gamma}{2}\right)>0\;\;\;;l_1\left(\epsilon+\dfrac{\Gamma}{2}\right)>l_2\;\;\;;l_1l_2\left(\epsilon+\dfrac{\Gamma}{2}\right)>\left(\epsilon+\dfrac{\Gamma}{2}\right)^2l_3+l_2^2
	\end{align}
	which takes the same form as represented in Ref.\cite{satya26} while discussion the effects of linear as well as quadratic coupling in an optomechanical system simultaneously.
	\section{Conclusion}\label{sec5}
	In this article we have studied a quadratically coupled optomechanical system and observed the variation of \textit{Transmission Intensity} $\mathcal{T}$ with coupling strength $g$ and $\Delta$. A brief discussion of the variation has been given in section (\ref{sec3}). Later, we have constructed \textit{Drift Matrix} by adiabatically eliminating the atomic degree of freedoms. Finally we have given the stability conditions of the quadratically coupled optomechanical system. In Fig.[\ref{fig:variation-with-delta}] at $\Delta_e=0$ we can observe the \textit{transmission amplitude} is minimum because, the pumping frequency is at resonance with the two level atom frequency; which amplifies the probability of absorption of incident photons and excitement of the atomic system. In Fig.[\ref{fig:variation-with-g}] we have discussed the variation of $\mathcal{T}$ with coupling between cavity mode and two level atomic mode $g$. As the coupling decreases, due to less absorption of photons by two level system we expect a better optical response as an output through $\mathcal{T}$. The variation of optical response with different parameter of the system definitely gives a better insight of optical response of a quadratically coupled optomechanical system; along with the construction of drift matrix we gave an enhanced insight about the effect of quantum noise on the system. In Fig.[\ref{fig:transmission-with-cavity-detuning}] we have reproduced the results discussed in Ref.\cite{farooq23} for quadratically coupled system.
	\section{Acknowledgements}
	A. Kundu would like to express his gratitude to Dr. R. Srikanth for fruitful discussions and comments on the draft. Moreover the author would like to thank U. Shrikant for his helpful reviews.
	
\end{document}